\documentclass[twocolumn,superscriptaddress
]{revtex4-2}

\usepackage{tikz} 
\usepackage{amsmath,amsfonts,amsthm, mathtools}
\usepackage{float}
\usepackage{dsfont}
\usepackage{csquotes}
\usepackage{amssymb}
\usepackage{verbatim}
\usepackage{mathtools}
\usepackage{listings}
\usepackage{xcolor}
\usepackage{caption}
\usepackage{dcolumn}
\usepackage{booktabs}
\lstdefinestyle{mypython}{
    language=Python,
    basicstyle=\ttfamily\small,
    keywordstyle=\color{blue}\bfseries,
    stringstyle=\color{green!50!black},
    commentstyle=\color{gray},
    numbers=left,
    numberstyle=\tiny,
    stepnumber=1,
    numbersep=5pt,
    breaklines=true,
    breakatwhitespace=false,
    showstringspaces=false,
    tabsize=4,
    captionpos=b
}
\usepackage{graphicx}
\usepackage{hyperref}
\usepackage{cleveref}
\usepackage{subcaption} 
\usepackage{rotating}
\usepackage{thmtools, thm-restate}
\usepackage{bm}
\usepackage{hyperref}
\usepackage{mathtools}
\DeclareUnicodeCharacter{2212}{-}
\usepackage{mathbbol}
\usepackage{pifont}
\usepackage{appendix}
\usepackage{tocloft}
\usepackage{titletoc}
\definecolor{darkred}  {rgb}{0.5,0,0}
\definecolor{darkblue} {rgb}{0,0,0.5}
\definecolor{darkgreen}{rgb}{0,0.5,0}
\hypersetup{
  colorlinks = true,
  urlcolor  = blue,         
  linkcolor = darkblue,     
  citecolor = darkgreen,    
  filecolor = darkred       
}

\usepackage{multirow}
\newcommand{\ket}[1]{|#1\rangle}
\newcommand{\bra}[1]{\langle #1|}

\usepackage{blkarray}
\usepackage{adjustbox}
\usepackage{graphicx}
\usepackage{dcolumn}
\usepackage{bm}
\usepackage[shortlabels]{enumitem}

\numberwithin{equation}{section}

\begin{document}

\title{A Universal Quantum Information Preserving Photonic Switch for Scalable Quantum Networks}
\author{Jiapeng Zhao}
\email{penzhao2@cisco.com}
\affiliation{Quantum Labs, Cisco Systems, 3232 Nebraska Ave, Santa Monica, California, 90404, USA}

\author{St\'ephane Vinet}
\thanks{These authors contributed equally to this work.}
\affiliation{Quantum Labs, Cisco Systems, 3232 Nebraska Ave, Santa Monica, California, 90404, USA}

\author{Amir Minoofar}
\thanks{These authors contributed equally to this work.}
\affiliation{Quantum Labs, Cisco Systems, 3232 Nebraska Ave, Santa Monica, California, 90404, USA}

\author{Michael Kilzer}
\affiliation{Quantum Labs, Cisco Systems, 3232 Nebraska Ave, Santa Monica, California, 90404, USA}

\author{Lucas Wang}
\affiliation{Department of Physics, University of California, Santa Barbara, California 93106, USA\\}

\author{Galan Moody}
\affiliation{Department of Electrical and Computer Engineering, University of California, Santa Barbara, California 93106, USA}

\author{Vijoy Pandey}
\affiliation{Quantum Labs, Cisco Systems, 3232 Nebraska Ave, Santa Monica, California, 90404, USA}

\author{Ramana Kompella}
\affiliation{Quantum Labs, Cisco Systems, 3232 Nebraska Ave, Santa Monica, California, 90404, USA}

\author{Reza Nejabati}
\affiliation{Quantum Labs, Cisco Systems, 3232 Nebraska Ave, Santa Monica, California, 90404, USA}

\date{\today}

\begin{abstract}

\noindent 

\noindent Quantum networks are a keystone of the quantum internet. However, existing implementations remain largely confined to static point-to-point links due to
the absence of a switching paradigm capable of dynamically routing fragile quantum entanglement without introducing decoherence.  Here, we propose the Universal Quantum Switch, a foundational building block allowing on-demand, non-blocking, and encoding-agnostic routing of quantum information, as well as seamless modality conversion between disparate quantum platforms. 
We develop a prototype in thin-film lithium niobate and experimentally demonstrate robust switching with $\le 4\%$ decoherence via thermo-optic modulation and high-speed electro-optic switching of arbitrary entangled states at 1 MHz. Moreover, we show that our platform can support reconfiguration speeds up to 1 GHz.
To our knowledge, this work represents the first demonstration of multi-node dynamic entanglement distribution at these speeds. Complementing these experimental results, we project the architecture's scalability, showing dimension-independent decoherence, and provide a scalable, interoperable building block for heterogeneous quantum network fabrics. 
\end{abstract}
\maketitle

\section{Introduction}
Quantum networks provide a scalable framework for interconnecting spatially separated quantum systems, enabling transformative applications such as distributed quantum computing \cite{PhysRevA.59.4249, aghaee2025scaling, main2025distributed}, distributed quantum sensing \cite{komar_quantum_2014, Gottesman, zang2024quantum, kim2024distributed}, and quantum cryptography \cite{BENNETT20147}. Despite significant progress in quantum networks over optical fiber and free-space channels, most existing implementations remain limited to static, point-to-point links between fixed node pairs \cite{wengerowskyentanglementbased2018, Joshi, chen_integrated_2021, van2022entangling, zhou2024long, craddock2026highratescalableentanglementswapping,Zhang2025}. Such network architectures fundamentally lack scalability and efficiency: resource requirements scale quadratically with network size, while the probabilistic nature of the entanglement distribution leads to poor utilization and idle resources \cite{PRXQuantum.2.040304, vinet_reconfigurable_2025,Vinet:26, appas_flexible_2021, zhao2025scalable}. However, transitioning toward scalable quantum network architectures requires dynamically routing quantum states across multiple nodes utilizing an integrated quantum switching platform while rigorously preserving quantum coherence, a capability that remains largely unaddressed. \par

The primary challenge comes from the fact that existing optical switching technologies, which underpin classical networks, are inherently incompatible with quantum information \cite{hall2011ultrafast, wang2025low, PW2026}. Photonic qubits are highly sensitive to impairments such as polarization drift, timing jitter, chromatic dispersion, and phase noise. These perturbations collectively introduce decoherence, leading to the loss of quantum information \cite{PW2026}. Most existing mitigation approaches, based on classical probing and pre-/post-compensation (before/after switch) \cite{craddock2026highratescalableentanglementswapping,Zhang2025, van2022entangling, zhou2024long}, are reactive. This leads to additional latency and overhead, and does not scale to dynamically reconfigurable networks. \par

These challenges in achieving scalable quantum networks are further amplified by the emerging heterogeneous vision of the quantum internet, where diverse quantum platforms interoperate across different encoding modalities within a unified network fabric \cite{kimble_quantum_2008,doi:10.1073/pnas.1419326112}. The successful distribution of quantum information relies on interoperable interfaces converting quantum states between distinct photonic degrees of freedom \cite{nehra2026photonic}. Achieving this modality conversion without disrupting the underlying quantum information remains a critical milestone that has yet to be demonstrated. \par

\begin{figure}
    \centering
    \includegraphics[width=\linewidth]{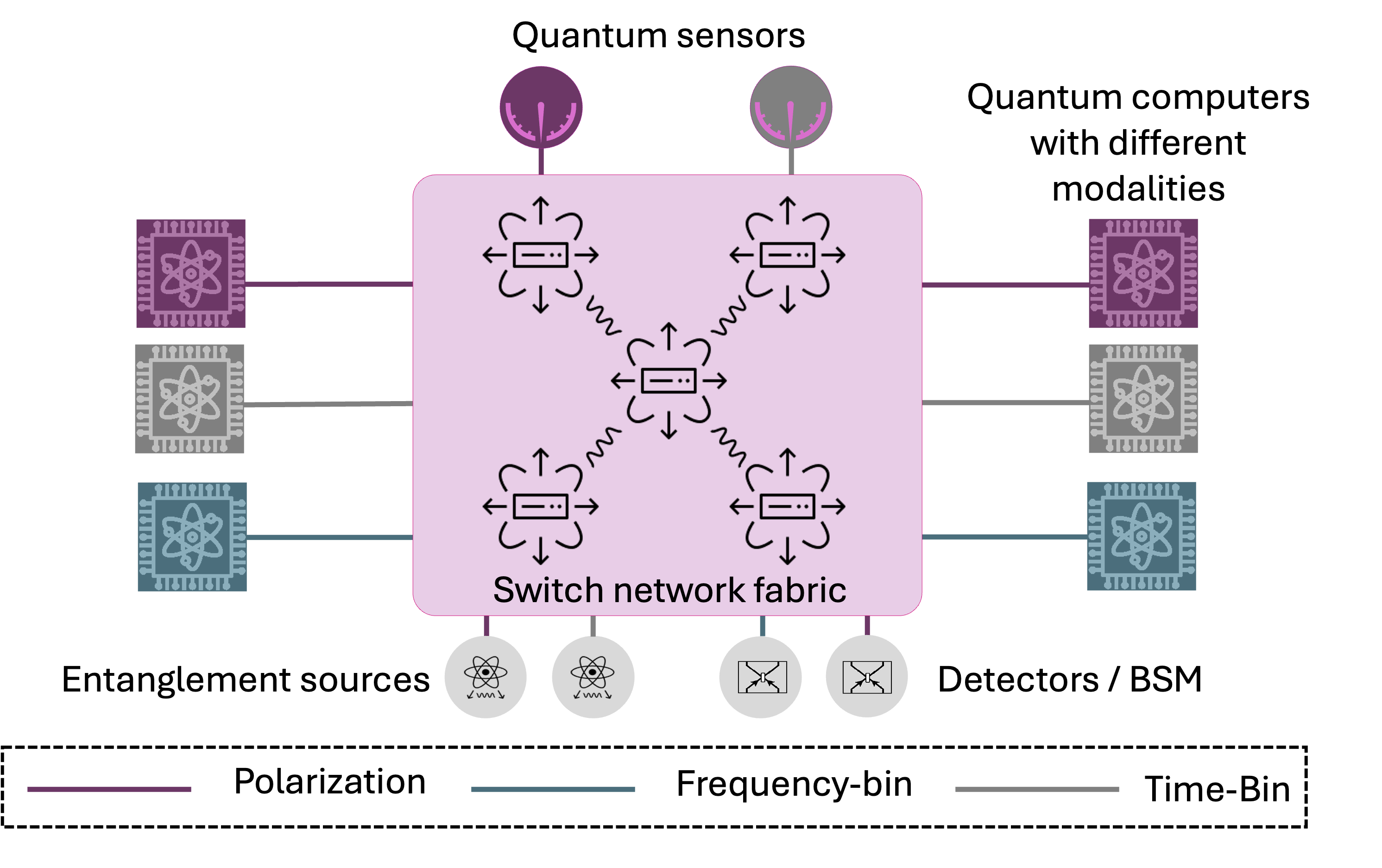}
    \caption{\textbf{Switched Quantum Network.} Conceptual quantum network centered around the quantum switch. The system ensures quantum state integrity and entanglement preservation while providing encoding-agnostic operation across diverse modalities. The switch supports time- and space-multiplexed utilization of shared critical resources whilst providing a scalable framework for the interconnection of quantum computers and quantum sensor. 
    }
    \label{fig:intro}
\end{figure}

To address this critical technical gap, we propose a Universal Quantum Switch (UQS), a foundational building block for scalable and dynamically reconfigurable quantum networks (shown in Fig.~\ref{fig:intro}). The proposed UQS enables on-demand, non-blocking, encoding-agnostic switching of quantum information while preserving quantum coherence. This allows seamless interoperability between heterogeneous quantum devices and facilitates efficient resource sharing \cite{shapourian2025quantumdatacenterinfrastructures}. By overcoming the limitations of conventional optical switches and enabling dynamic quantum state distribution, the UQS provides the essential functionality required for scalable quantum network architectures, including distributed quantum computing and entanglement-as-a-service.

The remainder of the paper is organized as follows. We first introduce the concept and architecture of the Universal Quantum Switch in Section~\ref{sec:architecture}. After this, we show a proof-of-concept demonstration of our architecture based on a thin-film lithium niobate (TFLN) switch specifically designed for polarization encoding. Section~\ref{sec:characterization} presents the device benchmarking results, including classical and quantum characterizations. In Section~\ref{sec:quantum}, we demonstrate entanglement-preserving switching for an arbitrary entangled state under both thermo- and electro-optic modulation. Finally, the scaling potential of our architecture is analyzed in Section~\ref{sec:discussion}.

\section{Universal Quantum Switch Architecture}\label{sec:architecture}

\begin{figure}
    \centering
    \includegraphics[width=\linewidth]{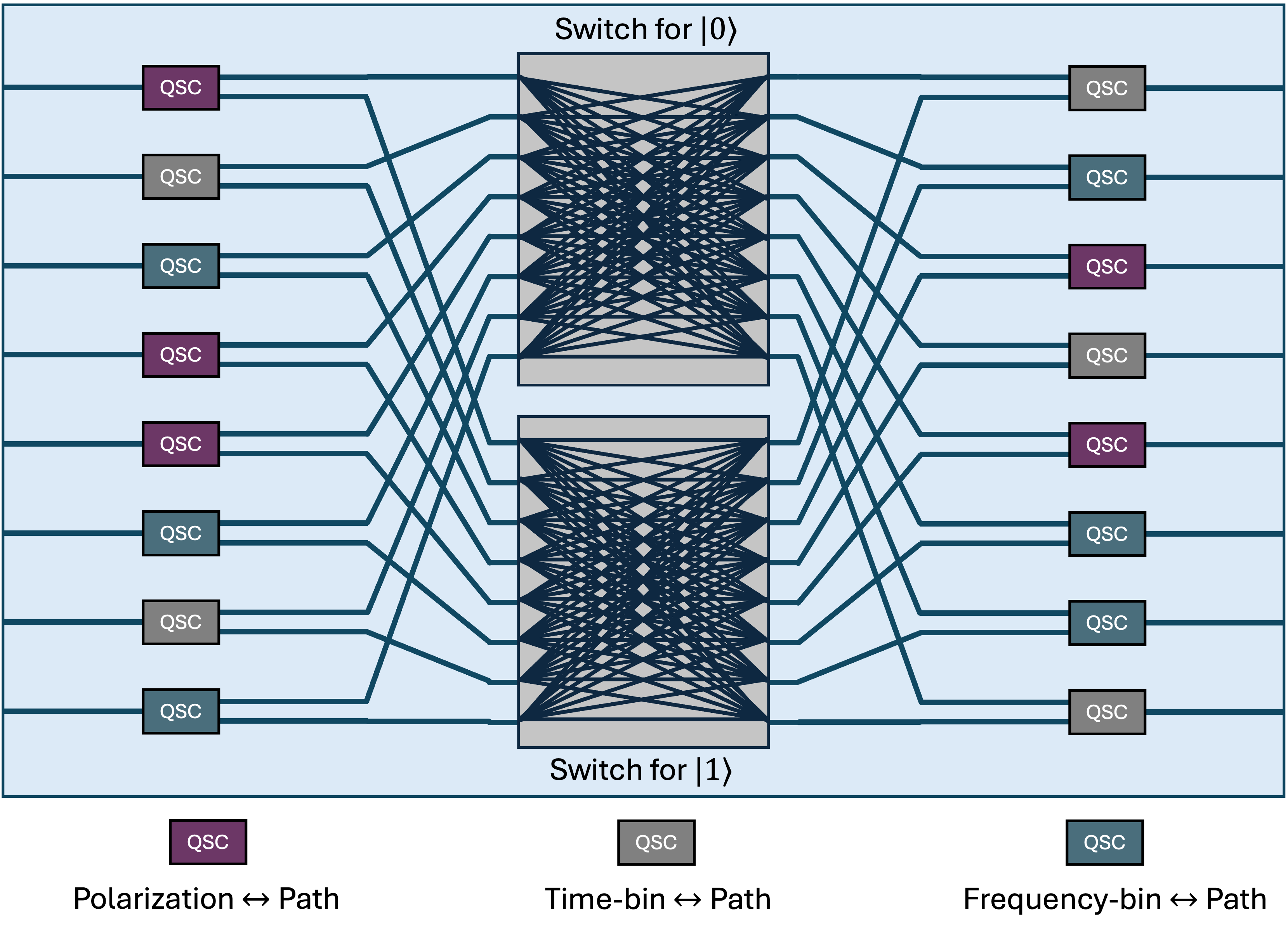}
    \caption{\textbf{Architecture of the Universal Quantum Switch.} The input quantum state converters (QSCs) enable conversion to path-encoding to ensure quantum information is routed through two identical photonic switches. The output QSCs convert the quantum information back to the desired output encoding modality. Both QSCs can be implemented in either an integrated or pluggable manner with an arbitrary combination of encoding modality.}  
    
    \label{fig:concept2}
\end{figure}

The architecture of the UQS is presented in Fig.~\ref{fig:concept2} \footnote{The architecture described herein is subject to a pending patent application from Cisco Systems, Inc.}. It consists of three functional stages to decouple quantum information from its physical carrier while ensuring flexible switching and encoding reconfiguration. These three main stages are as follows: (1) input quantum state converters (QSCs), (2) a non-blocking, all-to-all-connected photonic integrated switch, and (3) output QSCs. 

The input QSCs are responsible for mapping incoming quantum states from their native discrete-variable encoding—such as polarization, time-bin, or frequency-bin—into the path encoding used within the switch fabric. These conversions can be physically implemented via existing integrated photonics technologies, including but not limited to polarization rotator splitters, optical modulators, and quantum frequency converters \cite{ren2025chip, banic2024integrated, wright2017spectral, zhu2022spectral, singh2019quantum, lo2017precise}. Following the modality conversion, the logical $\ket{0}$ and $\ket{1}$ are routed to the upper and lower photonic switch modules, respectively.
The switch matrix is composed of a pair of non-blocking, all-to-all connected switches, which are designed to be indistinguishable. In this configuration, each switching element manipulates one logical bit of the path encoding, collectively enabling arbitrary switching while preserving the coherence of quantum states. \par
Following the photonic switch modules, both logical bits interfere at the output QSC to be encoded back into the desired modality. A key feature of this architecture is its reconfigurability, enabling flexible interconnection between heterogeneous quantum modalities. Specifically, it supports an arbitrary mapping between any input encoding and any output encoding through an appropriate selection of QSC modules at stages (1) and (3). This capability of quantum state conversion is fundamental for enabling interoperability across heterogeneous quantum systems and represents a critical requirement for scalable quantum networking. Moreover, since the routing of quantum information is based on two indistinguishable paths, the preservation of quantum coherence is passive, thus eliminating the need for classical probes and pre/post compensation  while enabling dynamical switching at high speed.

\section{Implementation and Validation}
\subsection{Device characterization and benchmarking}\label{sec:characterization}

\begin{figure*}[t!]
    \centering
    \includegraphics[width=\linewidth]{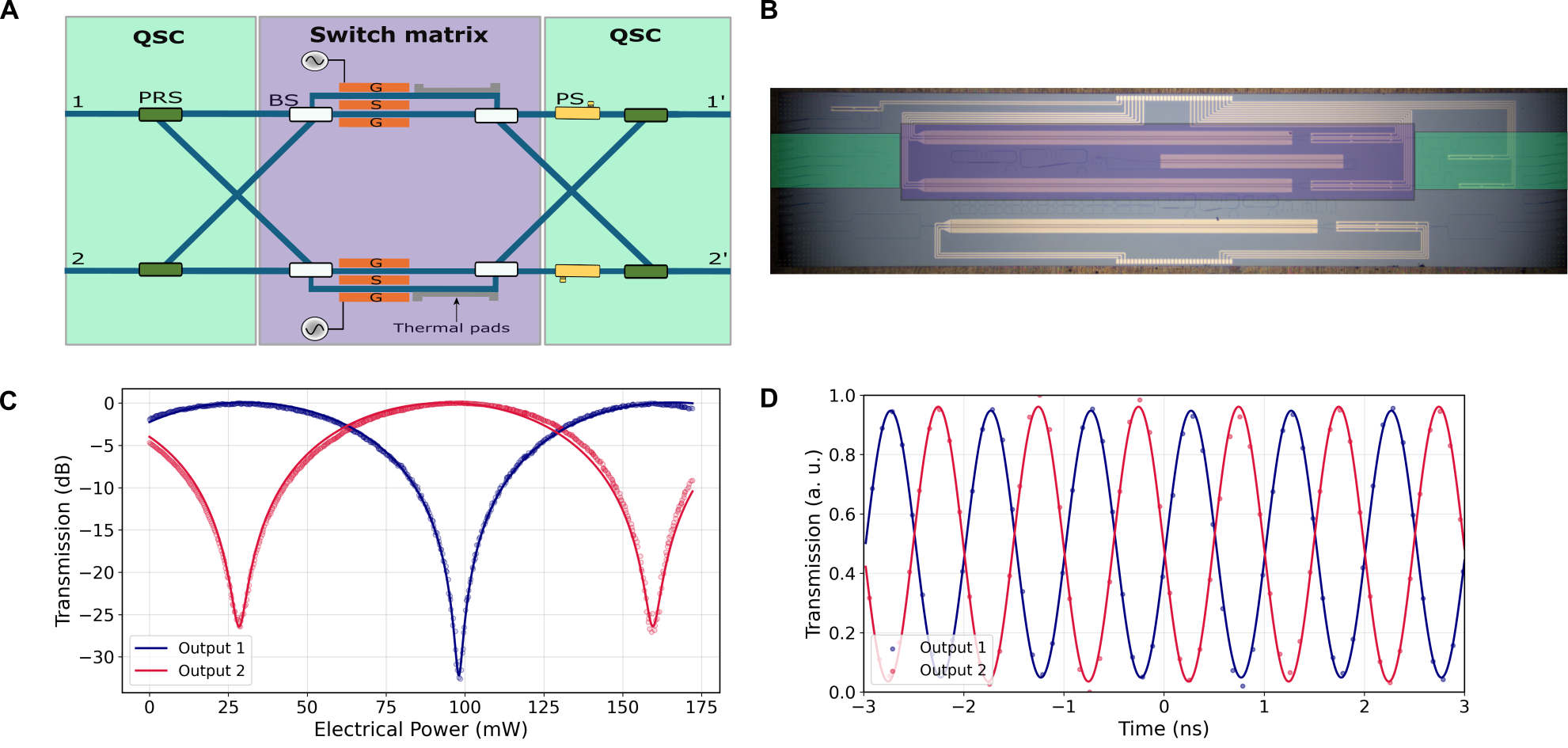}
    \caption{\textbf{Schematic of the quantum switch. (a)}
    Simplified sketch of the device for polarization encoding. 
    \textbf{(b)} TFLN photonic integrated circuit combining both QSCs and the switch matrix, respectively highlighted in green and purple. \textbf{(c)} Normalized optical power when varying the driving voltage of TO phase shifters to characterize the half-wave power and ER of MZI. \textbf{(d)} Fast switching between two output ports when driving the EO modulator with a sinusoidal waveform at 1 GHz rate to determine half-wave voltage. 
    }
    \label{fig:device}
\end{figure*}

In this work, we experimentally demonstrate the concept of the UQS based on a $2\times2$ TFLN photonic integrated circuit (PIC).
The $2\times2$ photonic switch serves as the fundamental building unit for high-dimensional integrated quantum switches \cite{psiquantum2025manufacturable, aghaee2025scaling, suzuki2020nonduplicate, zhao2023polarization, sun2019silicon, bartolucci2021switch}. By demonstrating quantum-coherence-preserving switching of this building block, we establish the viability of scaling this architecture to arbitrary $N$ dimensional quantum switch fabrics. The PIC architecture follows the switch architecture explained in section II. As shown in Fig.~\ref{fig:device}\textcolor{blue}{a}, the PIC utilizes QSCs, highlighted in green boxes, for polarization encoding at its input (polarization to path) and output (path to polarization), as well as two $2\times2$ integrated photonic switches marked in purple. We focus on switching polarization-encoded photons, as polarization is the most challenging degree of freedom to maintain loss and phase symmetry in PICs, to rigorously test the robustness of our architecture \cite{psiquantum2025manufacturable, aghaee2025scaling, suzuki2020nonduplicate, zhao2023polarization, sun2019silicon, bartolucci2021switch}.\par

In this design, QSCs are implemented using polarization rotator–splitters (PRSs) that route horizontally ($\ket{H}$) and vertically ($\ket{V}$) polarized photons to two nominally identical Mach-Zehnder interferometers (MZIs). The measured polarization extinction ratio (PER) and polarization-dependent loss (PDL) of the PRS, as well as the insertion loss (IL) of the chip are summarized in Table~\ref{table1}. The IL of the edge couplers is measured to be 1.87 dB per facet, in agreement with the design specification of $\sim$ 2 dB per facet \cite{Hyperlight2026}. Thus, the entire device loss excluding the coupling loss is 1.54 dB for output 1 and 1.43 dB for output 2. The IL, PER, and PDL performance at 1551.72 nm are consistent with state-of-the-art demonstrations \cite{song2023fully, wang2024polarization, zhao2026polarization}. \par

\begin{table}[ht!]
    \centering
    \begin{tabular}{|c|c|c|}
\hline
Metric & Output 1 (dB) & Output 2 (dB) \\
\hline
PER ($\ket{H}$) & 23.25                  & 19.36                  \\
PER ($\ket{V}$) & 24.79                  & 19.69                  \\
PDL             & 3.50                   & 3.30                   \\
Insertion Loss (IL)  & 5.25                   & 5.17                   \\ 
IL excluding coupling loss  & 1.54                   & 1.43                   \\ 
\hline
\end{tabular}
\caption{Measured performance metrics for the integrated TFLN quantum switch.}
\label{table1}
\end{table}

As shown in Fig.~\ref{fig:device}\textcolor{blue}{a} and \textcolor{blue}{b}, each MZI integrates two modulators: thermo-optic (TO) and electro-optic (EO) modulators. The EO modulator has a length of $10$ mm and the TO modulator has a length of $2.2$ mm. Slow switching ($\leq$ 5 kHz) can be implemented via TO modulators while fast switching ($\leq$ 1 GHz) is implemented by EO modulators. To ensure optimal performance during EO modulation, TO modulators are also used to establish and maintain the quadrature bias point. This dual-actuation strategy offers several advantages: it decouples the DC bias control from high-speed electrodes, simplifies the characterization process, and improves long-term stability by mitigating the effects of DC bias drift commonly associated with EO materials \cite{Wang:22, li_ultra-broadband_2026}. The performance of each TO modulator is characterized by injecting a pure polarization state aligned with its corresponding channel. As shown in Fig.~\ref{fig:device}\textcolor{blue}{c}, the measured transfer characteristics exhibit a half-wave power $P_\pi$ of $\sim 67.9$~mW and $\sim 65.5$~mW for TO modulators on two output ports with extinction ratios of 32.24 dB and 26.44 dB, respectively, aligning with current state-of-the-art performance \cite{assumpcao_thin_2024}. The lower extinction ratio at output 2 may originate from differences in the splitting ratios of the multi-mode interference (MMI) couplers. \par

\begin{figure*}[ht!]
    \centering
    \includegraphics[width=\linewidth]{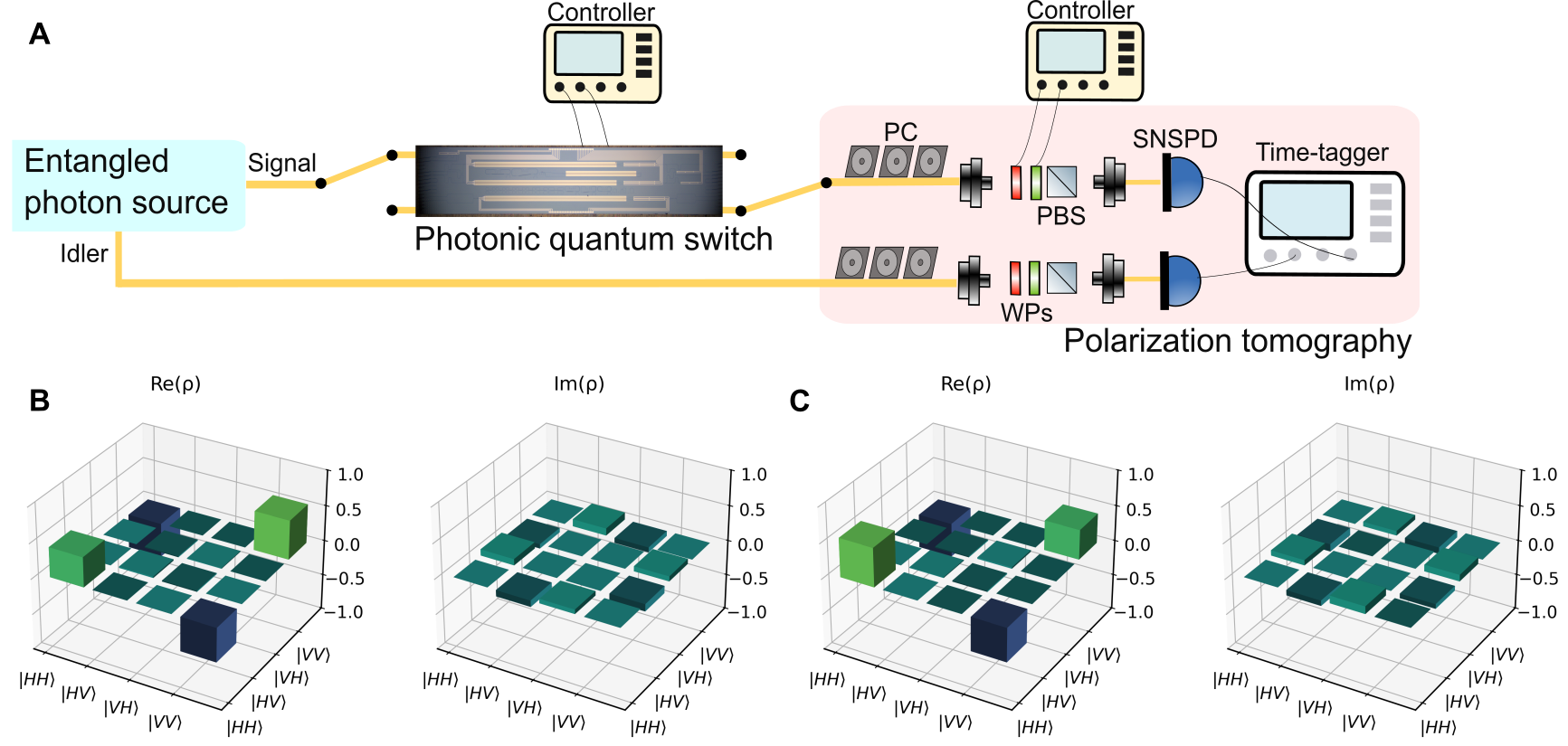}
    \caption{\textbf{Quantum state tomography (a)} Simplified sketch of the experimental setup for quantum characterization. An entangled photon source produces polarization-entangled photons at $1551.72$ $\rm{nm}$ (signal) and $1564.68$ $\rm{nm}$ (idler). The signal photon is routed through the UQS. After the PIC, both photons are sent to the polarization tomography system. Reconstructed density matrices for the input $\rho_{\rm{in}}$ \textbf{(b)} and output $\rho_{\rm{out}}$ \textbf{(c)} for connection $1\to 1$ (input $\to$ output ports). A fidelity $\mathcal{F}(\rho_{\rm{in}},\rho_{\rm{out}})=0.98$ is obtained with purity $Tr(\rho_{\rm{out}}^2)=1$.}
    \label{fig:quantum}
\end{figure*}

\par For fast switching, EO modulators are driven by RF waveforms to route the input light between output ports over time. We characterize the switch performance at 1 GHz as shown in Fig.~\ref{fig:device}\textcolor{blue}{d}. The amplitude of the sinusoidal waveform at 1 GHz is adjusted, and the peak-to-peak voltages of the sine waves at the switch output ports are measured. As such, the half-wave voltage $V_{\pi}$ on the function generator, which is fed to the RF probe, is measured to be $\sim$ 2.5 V at a 1 GHz frequency and 1.2 V at a 1 MHz frequency, which is used later in Section~\ref{sec:quantum}. More details of the electronic control and interfaces used for driving both TO and EO modulators are explained in Appendix~\ref{appendix:C}.\par

\par To benchmark the UQS performance for switching quantum photonic states, we perform quantum state tomography of a polarization-encoded entangled photon pair before and after one photon traverses the UQS to show the preservation of quantum information. A simplified version of the experimental setup is shown in Fig.~\ref{fig:quantum}\textcolor{blue}{a}. Polarization-entangled photon pairs are generated using a fiber-based Sagnac interferometer containing an AlGaAs microring resonator, which produces correlated pairs via spontaneous four-wave mixing \cite{PRXQuantum.2.010337}. Details of the entangled photon source are provided in Appendix~\ref{appendix:EPS}. The signal photon is routed through the UQS, while the idler photon proceeds directly to the quantum state tomography system (see Appendix~\ref{appendix:QST}). \par

We first prepare the $\ket{\Phi^-}$ state at the input of the UQS. More explicitly, we employ polarization rotators to perform a unitary basis alignment, ensuring the entanglement source basis is co-aligned with the basis of TFLN chip ($\ket{H} \rightarrow \ket{TE}, \ket{V} \rightarrow \ket{TM}$), where TE and TM represent the fundamental transverse electric and transverse magnetic modes of the TFLN PIC. By controlling the TO modulators, we route the $\ket{\Phi^-}$ state between two switch outputs. The tomography results for each switch matrix configuration are summarized in Table~\ref{tab:tomo} and shown in Figure \ref{fig:quantum}. For all connections, the reconstructed density matrices exhibit an average purity ($Tr(\rho_{\rm{out}}^2)$) above $99\%$ and an average Uhlmann fidelity to the input state, $\mathcal{F}(\rho_{in},\rho_{out})> 94\%$. We define the quantum decoherence from the UQS as the average loss of purity and fidelity compared to the input state, leading to an average penalty from the UQS below 4$\%$. One can notice that output 2 has a lower fidelity than output 1. This difference is primarily attributed to the low PER of the PRS and the ER of the MZI of output 2 as previously reported. Furthermore, to ensure both switch matrices are indistinguishable, one TO phase shifter is added before each output QSC for phase matching purpose. In the current demonstration, we observe that the phase at output 1 is typically matched automatically without the requirement of phase compensation. However, a constant phase is usually required for output 2, and the phase noise induced by power fluctuations in the TO phase shifter is observed at 0.04 rad. This noise also contributes to the decoherence at output 2. 
\begin{table}
    \centering
    \begin{tabular}{|c|c| c|c|}
    \hline
        Connection (input-output) & $Tr(\rho_{\rm{out}}^2)$ & $C$ & $\mathcal{F}(\rho_{\rm{in}}, \rho_{\rm{out}})$\\
        \hline
          $1\to1$& 1 &$0.99$& 0.98\\
          $1\to2$& 0.96 &$0.93$& 0.89\\
          $2\to 1$&$1$&$0.99$&$0.98$\\
       $2\to 2$&$1$&$0.93$&$0.92$\\
         \hline
    \end{tabular}
    \caption{\textbf{Benchmarking results.} The purity of the reconstructed density matrix $Tr(\rho_{\rm{out}}^2)$, its concurrence ($C$), and its fidelity ($\mathcal{F}(\rho_{\rm{out}}, \rho_{\rm{in}})$) to the input state $\rho_{\rm{in}}$ are characterized in Appendix~\ref{appendix:EPS}. The average decoherence penalty of the switch is $\leq 4\%$.
}
    \label{tab:tomo}
\end{table}

\subsection{Dynamic switching of arbitrary entangled states}\label{sec:quantum}
Since an ideal UQS can be considered as an identity operator $\Hat{I}$ on polarization entangled states, it can perform dynamic switching for maximally entangled states with arbitrary phases. In this section, we experimentally demonstrate that our architecture enables dynamic switching of arbitrary entangled states, using either TO or EO modulators, while preserving quantum coherence. \par

To experimentally demonstrate this capability, we modify the experimental configuration shown in Fig.~\ref{fig:quantum}\textcolor{blue}{a} as follows: we first prepare an arbitrary entangled state $\ket{\Phi^\theta}$, where
\begin{equation}\label{eq:state}
    \ket{\Phi^{\theta}} = \frac{1}{\sqrt{2}}(\ket{HH}+e^{i\theta}\ket{VV}),
\end{equation} 
and then remove the polarization controller before the UQS. After this modification, our system mimics a three-node entanglement distribution network without polarization tracking. The dynamic switching of entangled states without quantum decoherence in such a system has not been demonstrated yet, which we characterize below.\par

We first show the preservation of quantum coherence under TO switching. An arbitrary entangled state $\ket{\Phi^{\theta = -0.49\pi}}$ is prepared and sent to the UQS. By controlling the TO modulators for logical $\ket{0}$s and $\ket{1}$s, the quantum state is switched between output 1 and 2. After the UQS, photons are measured at the quantum state tomography system yielding a fidelity of $96\%$ and a purity above $99\%$ for both output ports.

\par
\begin{figure}
    \centering
    \includegraphics[width=\linewidth]{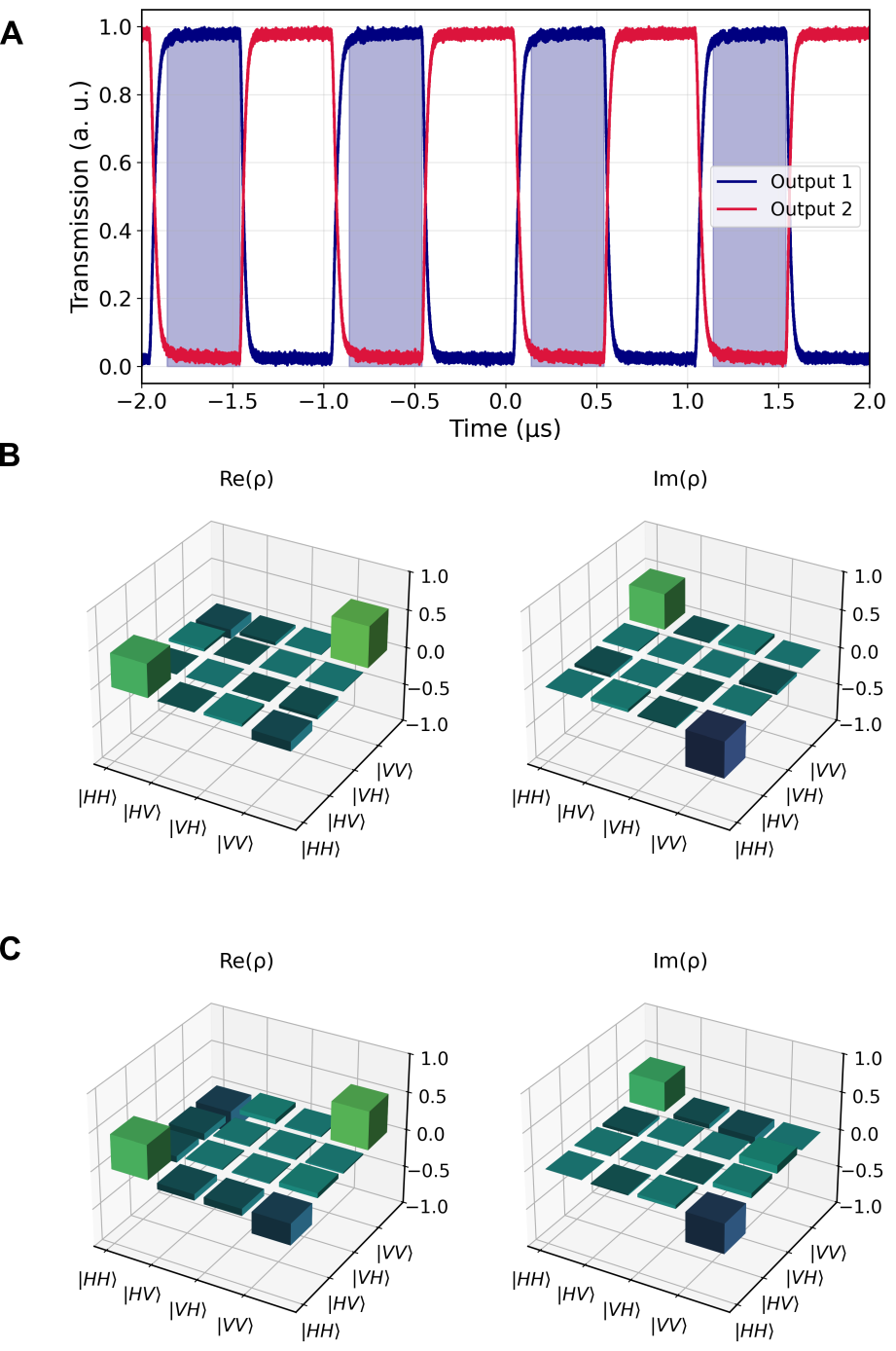}
    \caption{\textbf{Dynamic switching. (a)} Dynamic switching of the device when driving the EO modulator with a rectangular pulse at $1$ MHz. The gated section used for quantum state tomography is shown in the shaded region. Reconstructed density matrices for the input $\rho_{\rm{in}}$ \textbf{(b)} and output $\rho_{\rm{out}}$ \textbf{(c)} for connection $2\to 1$ (input $\to$ output ports). A fidelity $\mathcal{F}(\rho_{\rm{in}},\rho_{\rm{out}})=0.90$ is obtained with purity $Tr(\rho_{\rm{out}}^2)=1$.}
    \label{fig:dynamic}
\end{figure}

 Next, we demonstrate the preservation of quantum coherence under EO switching at 1 MHz. Another entangled state $\ket{\Phi^{\theta = -0.58\pi}}$ is prepared to show the flexibility of our architecture. As shown in Fig.~\ref{fig:dynamic}\textcolor{blue}{a}, two rectangular pulses with an amplitude of 1 V, a repetition rate of 1 MHz, and a $50\%$ duty cycle are sent to the corresponding RF pads. We observe that the settling time is $\sim$ 81.5 ns for each cycle, and thus for quantum state tomography, a 400 ns gated time window (with margin) is used during each data collection. Based on the tomography results, the UQS provides a fidelity $\mathcal{F}(\rho_{in},\rho_{out})\geq 90\%$. More details of the electronic control can be found in Appendix~\ref{appendix:C}. 

As one can notice, the quantum decoherence from the dynamic switching results is typically worse than the benchmarking results shown in Section~\ref{sec:characterization}. The major contribution of noise comes from the electronic control. Due to the drift in contact resistance between the probe and control pads on the chip, we observe a more severe fluctuation in the electrical signal applied on the switch when a high-speed signal is applied, resulting in a worse time-varying interference visibility. We expect electronic-photonic packaging of the chip will mitigate this issue.\par 

\section{Discussion}\label{sec:discussion}
We have validated the UQS architecture using an integrated $2\times2$ switch, which serves as the fundamental building block for arbitrary high-dimension switches\cite{psiquantum2025manufacturable, aghaee2025scaling, suzuki2020nonduplicate, zhao2023polarization, sun2019silicon, bartolucci2021switch}. In principle, the scalability of high-dimensional quantum switches is constrained by the accumulation of decoherence as the switch dimension $N$ increases. To quantify this, we utilize a theoretical model based on the weighted Pauli channel, which generalizes the standard depolarization channel to account for the specific noise contributions of the various integrated photonic components (see Appendix~\ref{app:model} for details). We find that the quantum decoherence in our architecture is mostly dimension independent, indicating a high level of scalability.

\par
\begin{figure}
    \centering
    \includegraphics[width=\linewidth]{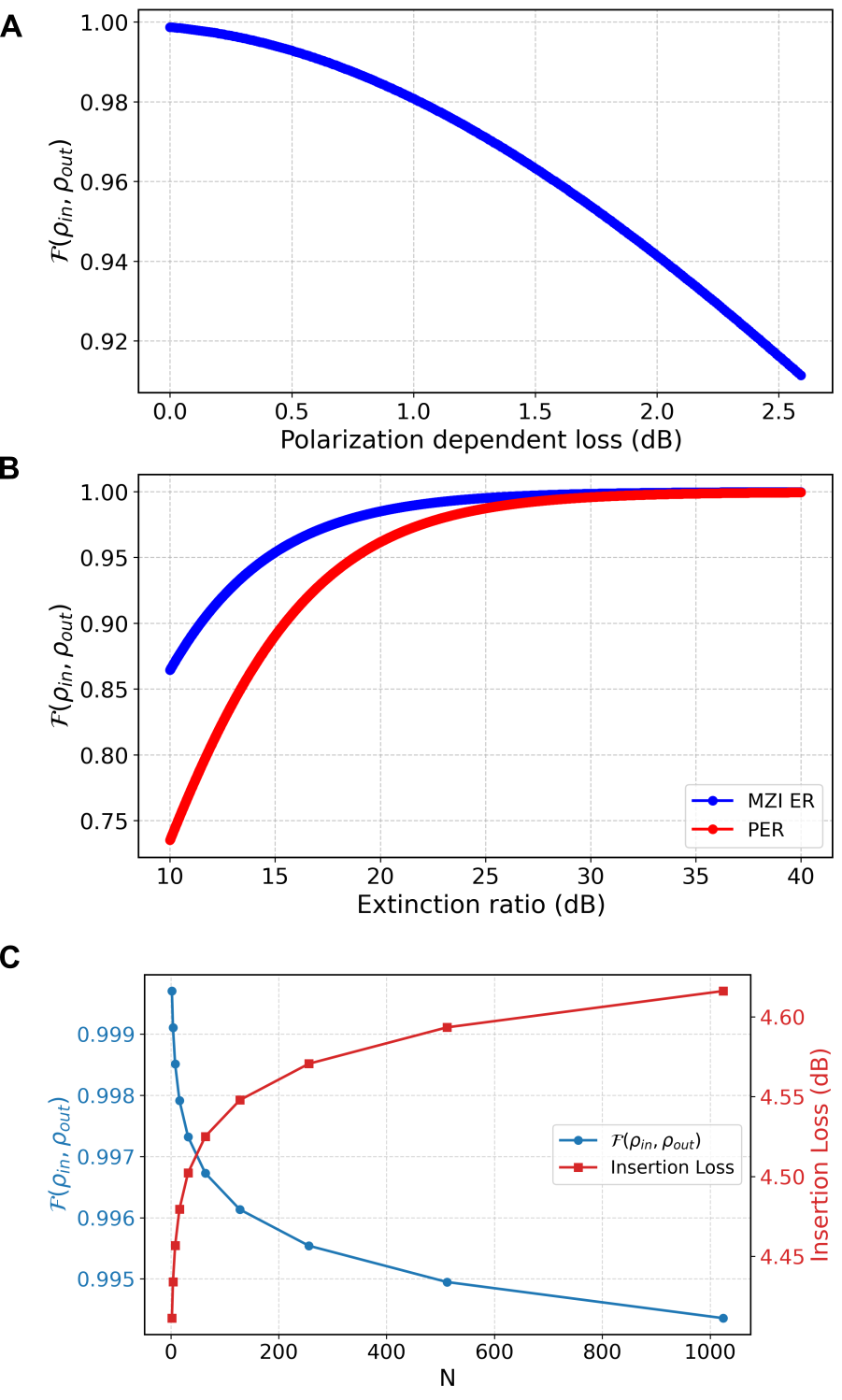}
    \caption{\textbf{Scaling potential of the UQS.} \textbf{(a)} UQS fidelity as a function of PDL per PRS. \textbf{b} UQS fidelity as a function of PER of PRS and ER of MZI. \textbf{c} UQS fidelity and IL as a function of dimension $N$. Note that the IL floor comes from the high coupling loss in the current chip. A low loss ($\leq 1$ dB) design can be implemented to reduce the IL floor to 2 dB.}
    \label{fig:sim}
\end{figure}

 The primary source of quantum decoherence comes from the limited performance in QSCs. In this context, the noise arises from the limited polarization extinction ratio and polarization dependent loss of the integrated PRSs. However, in our architecture, the number of QSCs for each photon is independent of the depth of the switching network. Consequently, the finite PER and PDL inherent to the PRS introduce noise that remains independent of the dimension of the switch $N$.
Furthermore, high performance is practical in the near term. As illustrated in Fig.~\ref{fig:sim}, maintaining a target fidelity $>99\%$ requires a PDL less than $0.64$ dB per PRS (Fig.\ref{fig:sim}a) and a PER  greater than $26.55$ dB (Fig.~\ref{fig:sim}b). These requirements are well within the performance benchmarks of recently reported designs \cite{zhao2026polarization}. The ER of the MZI also contributes to quantum decoherence but in a negligible way. With a measured ER of $32.24$ dB, the state fidelity can remain above 99.4$\%$ even for a $N = 1024$ switch, and with a projected ER at 35 dB, this fidelity can be improved to be above 99.7 $\%$. Therefore, the fidelity only drops slightly with an increasing $N$. \par

Maintaining path indistinguishability between arbitrary QSCs is essential for quantum-coherence preserving operation. In our current demonstration, we observe that the phase difference between arbitrary QSCs remains constant. Thus, the constructive interference between the logical $\ket{0}$ and $\ket{1}$ can be effectively guaranteed using static phase shifters. For a high-dimensional UQS, a look-up table to static phase shifters before the output QSC needs to be developed. Since each photon only travels through one phase shifter, the decoherence induced by phase noise does not increase with the dimension $N$. Regarding the path length symmetry, the propagation delay for logical $\ket{0}$ and $\ket{1}$ is precisely matched by the waveguide design. Consequently, no temporal walk-off is observed between these states, leading to high quantum interference visibility. Therefore, the quantum decoherence of the UQS can be considered as dimension independent.\par

Beyond decoherence, insertion loss represents a major bottleneck for high-dimensional photonic switches \cite{bartolucci2021switch, chen2023review, zhao2025scalable}. In a standard Bene\v{s} topology, the number of MZIs traversed by each photon scales as $\log_2 N$, resulting in substantial throughput loss as $N$ becomes large. In our current demonstration, the TFLN waveguides exhibit an insertion loss $\sim$ 0.2 dB/cm, leading to more than 8 dB loss for a $N=1024$ switch (see Fig.~\ref{fig:sim}\textcolor{blue}{c}). However, this can be significantly mitigated via advances in material and coupling technology. Indeed, by adopting TFLN platforms with a propagation loss of 1 dB/m \cite{zhu2024twenty, li2023ultra}, the waveguide-related loss can be reduced to 4.6 dB. Furthermore, while our current edge couplers impose a 3.74 dB loss floor (1.87 dB per facet), emerging designs enable coupling losses below 1 dB per facet with a PDL of 0.2 dB \cite{Hyperlight2026}.
Therefore, leveraging these recent experimental demonstrations, a projected device loss of 2.6 dB, with $N =1024$, is within reach with existing fabrication capabilities. \par

The current device is designed to be reconfigured at a speed of $\leq 1$ GHz, mainly because this rate can already satisfy the execution of most distributed quantum computing algorithms in major quantum computing platforms \cite{zhao2025scalable, wintersperger2023neutral, moses2023race, kjaergaard2020superconducting}. However, it is worth noting that the EO modulators on TFLN can be operated at speeds exceeding 100 GHz \cite{valdez2023100}, and our architecture does not have a limit on reconfiguration speed as the preservation of quantum coherence is independent of the active switch control.\par


Despite that our current experimental demonstration focuses on a fully-integrated solution, the flexibility of our architecture allows a modular implementation as well. Each type of QSC can be prepared in separate chiplets, and the combination of different types of QSCs will enable the arbitrary encoding modality conversion \cite{psaila2023detachable}. This configuration allows the switch to be adapted as an interface to accommodate diverse network encoding requirements.\par

The conversion between time-bin entanglement and path entanglement can be achieved using high-speed, low-jitter $1\times2$ switches and optical delay lines \cite{ren2025chip}. Wavelength multiplexer/demultiplexer can convert frequency-bins towards path-entanglement and vice versa, while quantum frequency conversion enables the erasure or creation of which frequency information \cite{singh2019quantum,wang_quantum_2023}.
Therefore, the demonstration of quantum-coherence-preserving dynamic switching with arbitrary encoding modality conversion is feasible and will be included in our next fabrication run.\par

The architecture shown in Section~\ref{sec:architecture} focuses on the discrete qubit encoding. However, the design is not restricted to two-level systems. Scaling up to high-dimensional encodings, i.e. qudits, is feasible by adding more integrated photonic switch modules at the stage (2), and corresponding paths to connect QSCs. Additional phase shifters on each path is also required to assure the constructive interference at the second QSC. However, as discussed above, these are static phase corrections that requires a high-dimensional look-up table. Meanwhile, the decoherence is still dimension independent as only two QSCs are required for each photon.\par

\section{Conclusion}
We propose a Universal Quantum Switch architecture that allows dynamic switching of arbitrary quantum states and encoding modality conversion without quantum decoherence. An experimental demonstration, based on integrated TFLN chips specifically designed for polarization encoding, is presented with an average Uhlmann fidelity $>94\%$, and an average purity $>99\%$. We demonstrate robust switching of arbitrary entangled states with a quantum decoherence $\leq4\%$ using slow, thermo-optic operation, and $\leq5\%$ using fast, electro-optic operation. To the best of our knowledge, the UQS is the first experimental demonstration of routing arbitrary entangled states at MHz rate without sacrificing quantum coherence. Based on the theoretical model of the UQS, our architecture only introduces additional quantum decoherence in a negligible amount with an increasing dimension $N$. Therefore, it is feasible to scale up the UQS at a minimal decoherence and insertion loss, which makes the UQS a fundamental building block for scalable and dynamically reconfigurable heterogeneous quantum networks.

\section{Acknowledgements}
The authors would like to thank Nathan Liu, Kevin Luke and Marko Lon\v{c}ar from Hyperlight Corporation for helpful discussions, as well as the design and fabrication of the TFLN chips.

\section{Data availability}
The data sets used and analyzed in this study are available from the corresponding author upon reasonable request.


\appendix
\section{Entangled photon source}\label{appendix:EPS}
\begin{figure*}[ht]
    \centering
    \includegraphics[width=\linewidth]{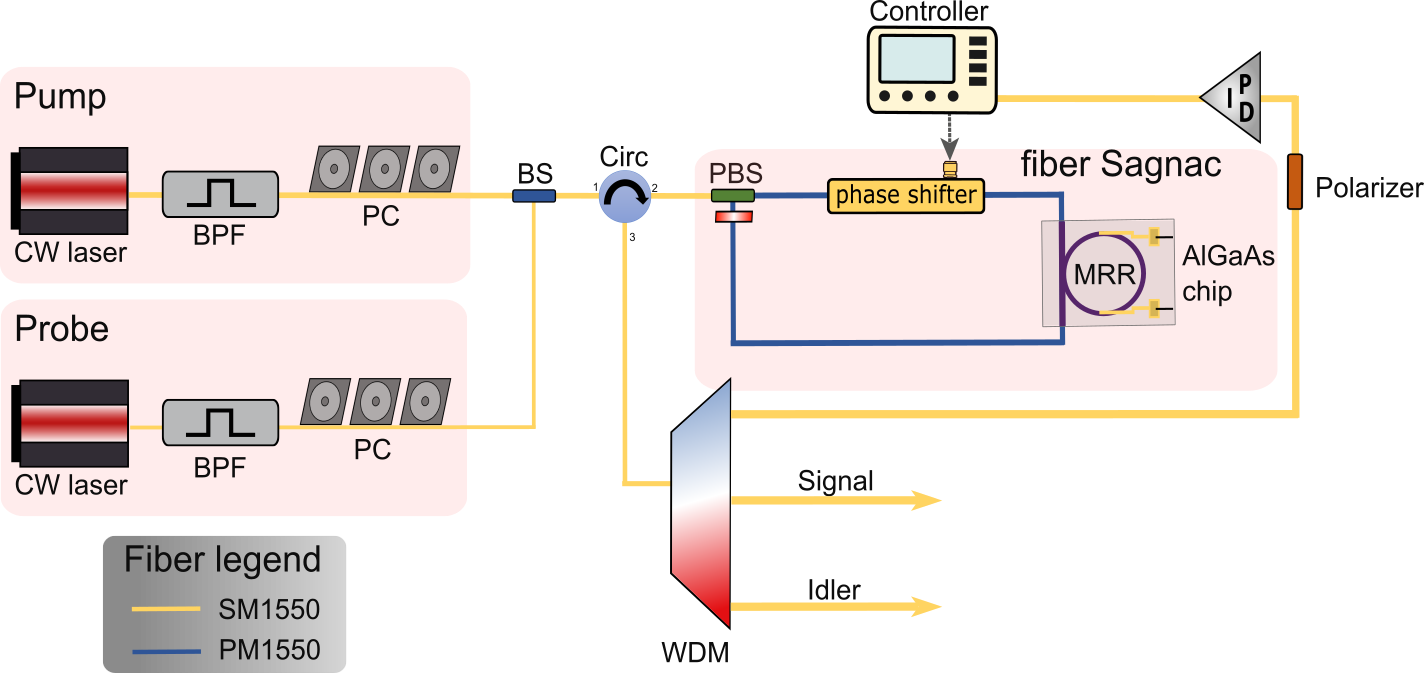}
    \caption{\textbf{Polarization entangled photon source.} Polarization entangled photons are generated in a fiber-based Sagnac interferometer via spontaneous four-wave mixing (SFWM) in a AlGaAs chip. The interferometer is actively locked using a probe laser with a servo controller.}
    \label{fig:source}
\end{figure*}

A detailed schematic of the entangled photon source is provided in Fig.~\ref{fig:source}\textcolor{blue}{a}.
The light from a tunable and continuous wave laser (\textit{Keysight N7778C}) is coupled to a fiber-based Sagnac interferometer. A polarization controller at the input of the Sagnac loop sets the pump polarization to the anti-diagonal polarization state $\ket{A}$ to enable bidirectional excitation of the resonator. The AlGaAs micro-ring resonator has a loaded $Q\sim 2\times 10^5$ with a free spectral range of $400$ GHz. The counter-propagating pump fields generate photon pairs in indistinguishable paths, whose superposition results in polarization-entangled states at the interferometer output. To ensure long-term phase stability of the interferometer, an auxiliary probe laser at $\lambda_{\rm{aux}} =1542.14$ $\rm{nm}$ is co-propagated through the Sagnac loop and used for active phase locking. The interference signal of the probe is monitored and fed back to a fiber phase shifter driven by a servo controller (\textit{NewFocus LB1005}), allowing stabilization of the relative phase between the counter-propagating paths. The locking point can be adjusted to control the phase of the entangled state $\ket{\Phi^\theta}$. Following the Sagnac loop, the signal and idler photons pass through an optical circulator and are then separated using an ITU-grid demultiplexer (\textit{Fiberdyne Labs}). The source system exhibits a loss of $23.7$ and $22.7$ dB for signal and idler photons, respectively. These losses were primarily attributed to source packaging and ITU grid demultiplexer.\par

\section{Quantum State Tomography}\label{appendix:QST}
To reconstruct the density matrix of entangled states, we perform quantum state tomography on the entangled photons directly emitted from the source, and then on photons after the UQS.\par

The quantum state tomography consists of motorized polarization analyzers, each comprising free-space collimators, a quarter-wave plate, a half-wave plate, and a polarizing beam splitter, which are all broadband devices to cover the entire telecommunication O and C bands. The tomography system has optical losses of $2.9$ and $4.3$ dB for signal and idler photons, respectively. After the collimators, both photons are detected with superconducting nanowire single-photon detectors (SNSPDs), and coincidence events are registered using a time-tagger (\textit{Swabian instruments Time Tagger X}). The SNSPD (\textit{ID Quantique}) has a quantum efficiency of $\geq 85\%$ and a timing resolution of $\leq 30$ ps. \par

Using a standard maximum likelihood estimation, we reconstruct the density matrix based on correlation measurement results. As shown in Fig.~\ref{fig:quantum}\textcolor{blue}{b}, the density matrix of the source, $\rho_{\rm{in}}$, is reconstructed. We obtain an Uhlmann fidelity $\mathcal{F}(\rho_{\rm{in}},\ket{\Phi^-}\bra{\Phi^-})=0.97$  to the $\ket{\Phi^-}$ Bell state, with purity $\rm{Tr}(\rho_{\rm{in}}^2)=1$ and concurrence $C=0.99$.\par


\section{Electronic interfaces}\label{appendix:C}
To characterize the TO modulation, the corresponding DC pads connected to the phase shifters are driven using a multi-contact wedge (MCW) probe with 32 Tungsten needles at $100$ $\mu$m pitch. We note that the measured resistance of the phase shifters placed in the TO modulators is $\sim 600 \Omega$ and
that of the ones placed in the output QSCs is $\sim 300 \Omega$. These values
match the design specifications since the length of the phase shifter in the TO modulator is twice that of the other. All measurements with the DC probe are performed under constant current operation. We observe a fluctuation in the measured contact resistance, leading to phase noises on the TO phase shifters as reported in Sec.~\ref{sec:characterization}.\par

 To explain the hardware devices used to drive the TO or EO modulators in detail, we divide the electronic control system into two operation regimes. To test dynamic switching at frequencies $<50$ MHz, a function generator (\textit{Keysight 33622A}) is used to generate two phase-synchronized square waves. The function generator interfaces with the PIC using a DC probe. Thus, the MCW is placed on both (i) ground-state-ground (GSG) pads of the EO modulator to drive the RF signal and (ii) DC pads associated with TO heaters to bias the modulator at the quadrature point for optimal performance. The optical signals at the switch output ports are detected using a high-gain photoreceiver with a transimpedance gain of 100 V/A and bandwidth of 10 MHz. Subsequently, the electrical signals are captured using a digital scope (\textit{Teledyne LeCroy WavePro 404HD}) having a BW of 4 GHz per channel and a sampling rate of 20 GSa/s. 

The combination of the EO modulator, MCW, breakout printed circuit boards (PCBs), ribbon cable connecting the PCBs, jumper wires, and terminal block adapter PCB forms a resistor, inductor, and capacitor (RLC) circuit. The RLC circuit is mismatched to the 50 $\Omega$ RF signal generator output impedance, leading to oscillations in the switch outputs at different frequencies. For switching the quantum signal, it is important that the square wave used to drive the phase modulator exhibits a flat amplitude response. To mitigate these oscillations, the optimal resistance value is determined by sending a 1 MHz rectangular waveform and processing the output amplitudes. The response can be modeled as an exponentially-decaying oscillatory system \cite{ogata2009, horowitz2015}. The exponential decay rate of the oscillation peaks is found via logarithmic decrement method. The main oscillation frequency ($f_\text{ring} \sim 15$ MHz) is found by applying a fast Fourier transform (FFT) on the rising edge of the normalized data. Consequently, the estimated values of inductance ($L$) and capacitance ($C$) are $L \sim 2702$ nH and $C \sim 41.2$ pF. To have the lowest 2\% settling time, a damping ratio of $\zeta \approx 0.7$ is chosen. Consequently, the total resistance required is $R_\text{total} = 2\zeta \sqrt{L/C} \sim 358\,\Omega$ \cite{ogata2009, horowitz2015}. Taking into account the source impedance of the function generator, the additional series resistance is $R_\text{series} = R_\text{total} - Z_\text{source} \sim 308\,\Omega$. Thus, a 330 $\Omega$ resistor is placed in series with each EO modulator's signal pad. This resulted in dampening the oscillations ($\zeta \sim  0.74$) and observing a settling time of $\sim$ 81.5 ns.

To test the dynamic switching performance at higher frequencies (
up to $1$ GHz), we use an arbitrary waveform generator (AWG) (\textit{Keysight M8190A}) with a sampling rate of 12 GSa/s and an analog bandwidth of 3.5 GHz. The AWG generates a sinusoidal signal at 1 GHz frequency with an amplitude of 200 mV. This signal is then amplified with a low-noise RF amplifier having a gain of 22 dB and noise figure of 1 dB at 1 GHz. This amplified RF signal is interfaced with the PIC using a customized RF probe with dual input channels and a pitch size of 100 $100$ $\mu$m with 6 needles placed on the GSGGSG pads on the PIC. Note that to optimize the performance of modulator, we also placed the MCW DC probe on the other pads connected to the TO heaters to bias at the quadrature point. The optical switch output ports are detected with two photodiodes (PD) (\textit{Thorlabs DET08CFC}) having bandwidth of 5 GHz and responsivity of $\sim$ 1 A/W. The output of each PD is terminated with a 50 $\Omega$ load resistance and connected to the sampling scope for data recovery.

\section{Theoretical model}\label{app:model}
The theoretical model of the UQS is based on a weighted Pauli channel, which generalizes the depolarization channel by accounting for specific hardware-induced decoherence including limited PER, PDL, the ER of MZI, and phase noise from the thermal phase compensator. Path mismatch between logical $\ket{0}$s and $\ket{1}$s is not included as per our experimental observation, this noise is much smaller than the phase noise induced by the thermal phase shifter.\par

In the theoretical model, we assume the input is an ideal polarization-entangled Bell state $\lvert \Phi^- \rangle$, represented by the density matrix $\rho_{in}$:
\begin{equation}
\lvert \Phi^- \rangle = \frac{1}{\sqrt{2}} (\lvert 00 \rangle - \lvert 11 \rangle) \implies \rho_{in} = \lvert \Phi^- \rangle \langle \Phi^- \rvert
\end{equation}

In the first step, we propagate the density matrix of the input state through the first PRS with a limited IL, PER and PDL. The physical impairments of the PRS are modeled using $2 \times 2$ Jones matrices applied to logical $\ket{0}$s and $\ket{1}$s.
We represent the coupling loss ($\eta_C$), PDL ($\eta_{PDL}$) with $J_{loss}$:
    \begin{equation}
    J_{loss} = \begin{pmatrix} \sqrt{\eta_{C} \eta_{PRS}} & 0 \\ 0 & \sqrt{\eta_{C} \eta_{PRS} \eta_{PDL}} \end{pmatrix},
    \end{equation}
and represent the PER induced crosstalk ($\epsilon_i$) with $J_{i,leak}$:
    \begin{equation}
    J_{i,leak} = \begin{pmatrix} \sqrt{1-\epsilon_i^2} & \epsilon_i \\ \epsilon_i & \sqrt{1-\epsilon_i^2} \end{pmatrix}.
    \end{equation}
where the subindex $i = 0, 1$ represents two logical qubits.

The switch depth $D$, the total number of MZIs that a photon travels through, is defined by the dimension of the switch $N$ as $D = 2 \log_2 N- 1$. After the MZI matrix, all logical $\ket{1}$s travel through a phase shifter with phase noise $\phi$. With efficiency amplitude $a = \sqrt{\eta_{mzi}^D}$ and random phase noise $\phi$, the Jones matrix of the switch matrix and phase compensator is:
    \begin{equation}
    J_{MZI}(\phi) = \begin{pmatrix} a & 0 \\ 0 & a e^{i\phi} \end{pmatrix}.
    \end{equation}
Note that, after first the PRS, both paths have TE mode only, and hence share the same efficiency.\par

Due to the finite extinction ratio of the MZIs, logical bits can be routed to other ports. The probability of a single logical bit successfully reaching the output is $p_s = (1 - ER)^D$, and the ER of each MZI, in both logical $\ket{0}$ and $\ket{1}$ switch matrices, is assumed to be identical. The output state is a weighted sum of three physical outcomes:
\begin{enumerate}
    \item Both logical $\ket{0}$ and $\ket{1}$ succeed: $P_1 = p_s^2$.\\
    The corresponding Jones matrix is: $J_{both} = J_{loss} J_{leak}J_{MZI}(\phi)J_{leak}J_{loss}$.
    \item Logical $\ket{0}$ only: $P_2 = p_s(1-p_s)$. \\
    The corresponding Jones matrix is: $J_{0} = J_{loss} J_{0,leak}J_{0,MZI}J_{0,leak}J_{loss}$.
    \item Logical $\ket{0}$ only: $P_3 = p_s(1-p_s)$. \\
    The corresponding Jones matrix is: $J_{1} = J_{loss} J_{1,leak}J_{1,MZI}(\phi)J_{1,leak}J_{loss}$.
\end{enumerate}
The final density matrix $\rho_{out}$ is calculated via a Monte Carlo simulation over $M$ iterations:
\begin{equation}
\rho_{out} = \frac{1}{M} \sum_{j=1}^{M} \left[ P_1 \rho_{both}(\phi_j) + P_2 \rho_{0} + P_3 \rho_{1} \right]
\end{equation}
where $\rho_k = (J_k \otimes I) \rho_{in} (J_k \otimes I)^\dagger$. Here, we assume that the $ER\ll1$ and the photon flux for each input port has the same level of sparsity. Therefore, we can ignore the probability that a photon from another input port is routed to the target output port. When part of the input ports have significantly larger flux compared to other ports, this assumption may not hold.\par

Based on the final state, we calculate the fidelity $F = \langle \Phi^- \rvert \hat{\rho}_{out} \lvert \Phi^- \rangle$, and the purity $P = \text{Tr}(\hat{\rho}_{out}^2)$ as a function of the PER, PDL, and ER of the MZI.

\bibliography{bib}
\end{document}